\documentstyle[12pt]{article}
\begin{document}
\baselineskip=1.2\baselineskip

\def\bea{\begin{eqnarray}}
\def\eea{\end{eqnarray}}

\renewcommand{\thefootnote}{\fnsymbol{footnote}}
\setcounter{equation}{0}
\setcounter{section}{0}
\renewcommand{\thesection}{\arabic{section}}
\renewcommand{\theequation}{\thesection.\arabic{equation}}

\pagestyle{plain}
\begin{titlepage}
\hfill{IP/BBSR/95-50}
\begin{center}
{\large{\bf {Spin polarisation by external magnetic fields, Aharonov-Bohm \\}
{flux strings, and chiral symmetry breaking in QED$_{3}.$}}}\par
\end{center}
\vskip 1.2cm
\begin{center}
{ Rajesh R. Parwani \\
Institute of Physics, Bhubaneswar-751 005,\\
INDIA.\footnote{e-mail: parwani@iopb.ernet.in }}
\end{center}
\vskip 2.0cm

\centerline{ May 1995}
\vskip 3.0 cm
\centerline{\bf Abstract}
{ In the first part, the induced vacuum spin
around an Aharonov-Bohm flux string in
massless three-dimensional QED is computed
explicitly
and the result is shown to agree with a general index theorem.
A previous observation in the literature,
that the presence of induced vacuum
quantum numbers which are
not periodic in the flux make an integral-flux AB string visible,
is reinforced.
In the second part, a recent discussion of
chiral symmetry breaking by external magnetic fields in
parity invariant QED$_{3}$ and its relation to the induced spin
in parity non-invariant QED$_{3}$ is  further elaborated.
Finally, other vacuum polarisation effects around
flux tubes in different variants of QED, in
three and four dimensions
are mentioned.}
\end{titlepage}

\newpage

\section{Introduction}

Three dimensional quantum electrodynamics (QED$_3$), with massive
electrons in an irreducible two-dimensional representation,
is peculiar compared to QED$_4$
in that it is
not invariant under a discrete parity transformation.
This breaking of parity by the
massive fermions has a striking consequence
that radiative corrections to the photon propagator generate an additional
gauge-invariant but parity-odd contribution to the effective lagrangian
for the gauge fields \cite{NSR}.
The last is the celebrated Chern-Simons term \cite{CS}.

The two-component nature of the fermions, the breaking of parity by fermion
masses, and the Chern-Simons term, form a triad of interlinked facts
which are responsible for various interesting effects in QED$_3$, one of
which  is the induction of fractional charge in the presence of a background
magnetic field \cite{NSR,rev}. To fix the notation, consider the lagrangian
of three-dimensional QED,
\begin{eqnarray}
{\cal{L}_{+}} &=&
{\bar{\Psi}}_{+} \left( i D_\mu \ \gamma^{\mu} -
 \ m \right)
\Psi_{+} \ ,
\label{lag1}
\end{eqnarray}
where $D_\mu = \partial_\mu + ie A_\mu$,
$\gamma^0 = \sigma^3, \ \gamma^1 =i\sigma^1, \ \gamma^2 = i\sigma^2$
and where $\Psi_+$ are two component spinors. (The subscript "$+$" in the
above equations has been inserted for latter convenience.)
As mentioned,
for $m \neq 0$,
${\cal{L_+}}$ is not invariant \cite{parity} under the  $2+1$ dimensional
parity operation $(x,y) \to (-x,y), \Psi_+ \to \sigma^1 \Psi_+$.
The mass parameter $m$ in (\ref{lag1}) is allowed to take
either sign $(\pm)$ so as to concisely describe the two inequivalent
irreducible representations of QED$_3$. If the potential $A_\mu$
decribes an external magnetic field with dimensionless flux
\bea
F &=& {e \over 2 \pi} \oint \vec{A} \cdot d\vec{r} , \nonumber
\eea
then the induced charge is given by
\bea
{\cal{Q}} &=& {e \over 2} \int d^2 x \ \langle \left[ {\bar{\Psi}}_{+},
\gamma^{0}
\Psi_{+} \right] \rangle  \label{q1} \label{c1}\\
&=& - {e \over 2} \ \int d^2 x \ \sum_{E} \mbox{sign}(E)
\psi_{E}^{\dag}(x) \psi_{E}(x) \label{q2} \label{c2}\\
&=& - \mbox{sign}(m){eF \over 2} \label{q3} \label{c3}  \ .
\eea
In Eq.(\ref{c2}), $\psi_E$ are eigenstates of the Dirac Hamiltonian $H_+$,
 and so
the induced charge is related  \cite{NSR,rev} to the spectral asymmetry
of $H_+$.
The last equation shows that in a time-independent external magnetic
field, the asymmetry in the spectrum is an invariant depending only on the
flux. There are numerous derivations of (\ref{c3}) in the literature for
generic magnetic field configurations \cite{NSR,rev,BB,Pol} and a
physical picture is as follows \cite{BB}:
free spinors have their spin correlated
with their energy, $s = {\mbox{sign}(mE) \over 2}$,
so that a perturbing magnetic
field, because of the spin-field interaction, polarises the virtual particle
states differently according to $E>0$ or $E<0$ thus causing an induced
charged cloud around the field.

There are other
induced quantum numbers in a time-independent external magnetic
field, like angular momentum \cite{MP}  $\cal{J}$
and spin \cite{BBY,Pol} $\cal{S}$
which depend only on the flux,
\bea
{\cal{J}} &=& - \mbox{sign}(m) \ { F^2 \over 4} \ , \label{ang}\\
{\cal{S}} &=& - \mbox{sign}(m) \ {|F| \over 4} \ , \;\;\; \mbox{for} \
m \to 0.
\label{spin}
\eea
All of the above induced quantities
depend on sign($m$) which indicates the crucial dependence
of the results on the physically different representations.
Also, as required, $\cal{Q}$ is odd under
charge conjugation (C) while $\cal{J}$ and $\cal{S}$ are C-even in agreement
with the transformation properties of the corresponding
single particle operators.
Actually, in explicit calculations, the correct C-symmetry is ensured
only if one computes properly
(a)symmetrised expressions and
is careful to choose regularisation schemes which
preserve the (a)symmetry. Note also the very important point that while
the expressions (\ref{c3}) and (\ref{ang}) for
$\cal{Q}$ and $\cal{J}$ are valid in the massive theory, $\cal{S}$ is an
invariant (depending only on the flux)
only in the massless limit of the theory.

An interesting external field configuration
is an Aharonov-Bohm \cite{AB} (AB)  string for which the magnetic field is
nonvanishing only at one point. Nevertheless, even in this singular case
the relations (\ref{c3}-\ref{ang})
are still valid as has been verified by explicit
calculations \cite{others,PG,FL,cone}. In
Ref.\cite{PG}
it was stressed that, because of the nonperiodic (in flux)
induced charge localised around
an AB string in $2+1$ dimensions, an otherwise
quantum mechanically unobservable
string can reveal its presence to a
probe electron in a {\it Gedanken} scattering experiment. This
may be seen as an example
in which quantum field (virtual many body) polarisation
effects invalidate
quantum mechanical expectations.

While
the induced charge and angular momentum have been computed for an
AB configuration, to my knowledge the induced spin has not. Other than
explicitly verifying the relation (\ref{spin})
for the singular AB configuration (see Sect.2),
and thereby reinforcing the last statement in the previous paragraph,
the calculation
involves a complication in that the limit $m \to 0$ in
(\ref{spin}) must be taken {\it before} doing the space-integral over
the spin-density. Actually, for an AB
configuration, the only free dimensionfull parameter in the problem is the
mass $m$ since the magnetic field enters only through the
dimensionless flux $F$.
Thus for the AB problem,
${\cal{T}} \equiv \int d^2 x \langle \Psi^{\dag}_{+} {\sigma^3 \over
2} \Psi_{+} \rangle$ is independent of the magnitude of $m$ and so trivially
${\cal{T}}^{AB} = {\cal{T}}^{AB}_{m \to 0}$. But since  ${\cal{T}}$
is {\it not} related in any simple
way to the spectral asymmetry, one does not
expect, in
general, its massless limit  to coincide with $\cal{S}$ (that is,
the massless limit required in (\ref{spin}) need not commute
with doing the space-integral), and this is discussed
further in Appendix A.

In $QED_4 $, if the electrons are massless, the theory has a chiral symmetry.
This is not true for
the lagrangian (\ref{lag1}) of $QED_3$ because there is no matrix
which anti-commutes with the $\gamma_\mu$'s. However by
enlarging the representation space of the spinors in
(\ref{lag1}), one can also
discuss chiral symmetry in massless $QED_3$. Consider thus
the three space-time
dimensional lagrangian
given by \cite{chiral}
\begin{equation}
{\cal{L}} = \bar{\Psi} \left( i D_\mu \ \Gamma^{\mu} - m \right)
\Psi \; ,
\label{lag2}
\end{equation}
where the {\it four}-component
spinor $\Psi$ forms a reducible representation of the Dirac
algebra
\begin{equation}
\Gamma^{0}=\pmatrix{\gamma^{0} & 0 \cr  0 & -\gamma^0} \ , \;\;\;
\Gamma^{1}=\pmatrix{\gamma^{1} & 0 \cr  0 & -\gamma^1} \ , \;\;\;
\Gamma^{2}=\pmatrix{\gamma^{2} & 0 \cr  0 & -\gamma^2} \ ,
\end{equation}
with $\gamma^0 = \sigma^3, \ \gamma^1 =i\sigma^1, \ \gamma^2
=i\sigma^2$ as before.
When $m \equiv 0$, this lagrangian
is invariant under a  $U(2)$ symmetry generated by $I, \Gamma^5,$
and $-i\Gamma^3$, where $\Gamma^3 \equiv \pmatrix{0&i \cr i &0}$
and $\Gamma^5 =i \Gamma^0 \Gamma^1 \Gamma^2 \Gamma^3$.
This symmetry is sometimes referred to as a flavour symmetry.
Alternatively, since the symmetry exists only for the massless
theory, one may also call it
a chiral symmetry.
Furthermore, by writing
$ \Psi \equiv \pmatrix{\Psi_{+} \cr \Psi_{-}}$, with $\Psi_\alpha
\;\ (\alpha = \pm )$
two-component spinors, the Lagrangian density (\ref{lag2})
is invariant (for any $m$) \cite{chiral}  under
the generalised
parity operation $(x,y) \to (-x,y), \ \Psi_{+} \to \sigma^1
\Psi_{-}$ and  $\Psi_{-} \to \sigma^1 \Psi_{+}$ .

As (\ref{lag2}) has
symmetries similar to those of physically interesting, but
technically difficult,
four-dimensional theories like QCD, it
has been used as a toy model to study such issues as
chiral symmetry breaking. Most of the last work has been on { \it dynamical}
chiral symmetry breaking, which involves the difficult problem of
finding self-consistent nonperturbative solutions to the
Schwinger-Dyson equations \cite{chiral}.
A simpler, but still informative, task is to
study chiral symmetry breaking induced by  { \it external}
 electromagnetic fields.
In Ref.\cite{GMS}
it was found that a space-time independent external { \it magnetic}
 field yielded
a nonzero value for the chiral order parameter, $\langle {\bar{\Psi}}_{+}
\Psi_{+} \rangle_{m \to 0}\ $. This result was generalised to inhomogeneous
(but still static) fields in Ref.\cite{RP}
by observing that in an external field (that is
ignoring virtual photon corrections), the chiral condensate of the lagrangian
$\cal{L}$ is related to the induced spin density of the
lagrangian $\cal{L}_{+}$.
In terms of two-component spinors,
Eq.(\ref{lag2}) may be written as
\begin{eqnarray}
{\cal{L}} &=& \sum_{\alpha =\pm} {\cal{L}_{\alpha}} \equiv
\sum_{\alpha =\pm}
{\bar{\Psi}}_{\alpha} \left( i D_\mu \ \gamma^{\mu} -
\alpha \ m \right)
\Psi_{\alpha} \ ,
\label{lag3}
\end{eqnarray}
so that
if $A_\mu$ is an external field,
(\ref{lag3}) describes two decoupled systems $\cal{L}_{\pm}$. The
fermion condensate is  therefore
\begin{eqnarray}
C(x;m) \equiv \langle\bar{\Psi}(x) \Psi(x) \rangle
&=& \sum_{\alpha = \pm} \alpha \ {\langle{\bar{\Psi}}_{\alpha}(x)
 \Psi_{\alpha}(x) \rangle}_{\alpha} \label{ord} \\
&\equiv& \sum_{\alpha= \pm } \alpha \ C_\alpha (x;m) \;
\label{cond} \\
&=& \sum_{\alpha = \pm} \alpha \ C_{+}(x; \alpha \ m)
\label{cond2} \, ,
\end{eqnarray}
where  ${\langle{\bar{\Psi}}_{+}
\Psi_{+} \rangle}_{+}$ denotes the expectation value of
${\bar{\Psi}}_+ \Psi_+$ in the $\cal{L_+}$ subsystem.
Thus in an external field
the properties of the
condensate $C(x; m \to 0)$ are determined completely by the
induced spin density ${1 \over 2} \ C_{+} = \langle {\Psi}^{\dag}_{+} {
\sigma^{3} \over 2} \Psi_{+} \rangle_{+}$
in the subsystem $\cal{L}_+$ which is a better studied problem.
An immediate consequence of (\ref{cond2}) is
$C(x; -m) = -C(x;m)$, so that the condensate $C(x;m \to 0)$
can be nonzero only if the massless limit is discontinuous, that
is it depends on $m \to 0^{\pm}$. That this is indeed the case
was noted in Ref.\cite{RP} and
Sect.(3) of this paper extends the discussion of chiral-symmetry
breaking given there.

The plan of the
paper is as follows. In Section 2.1 a derivation of
the relation (\ref{spin}) is
reviewed and in Sect. 2.2 it is verified by an explicit calculation in
the AB system. Section 3 repeats some of the discussion on chiral
symmetry breaking given in Ref.\cite{RP} and supplements it with
some qualitative comments. The
concluding section of this paper summarises some of the discussion
and highlights  other vacuum polarisation effects in
$QED_{3}$ and  $QED_{4}$. The
appendices contain some details of the calculations presented in
Sect.2. Appendix B also presents
a novel derivation of (\ref{c3}) for the AB
case in the massless limit. The reader is informed that for continuity
of discussion, some material from the short report \cite{RP} has been
reproduced in this article.

\setcounter{equation}{0}
\section{Induced Spin}

\subsection{General}
Consider the induced vacuum spin in the theory defined by (\ref{lag1}),
due to vacuum polarisation in a
time-independent external magnetic field,
\begin{eqnarray}
{1 \over 2} C_{+}(x;m\to 0) &=& {\lim_{m \to 0}} \,
\langle {\Psi}_{+}^{\dag} {\sigma^{3} \over 2}  \Psi_{+} \rangle \; .
\label{spin1}
\end{eqnarray}
Since $m$ can be of either sign, the limit in (\ref{spin1}) should
be understood correspondingly
as $m \to 0^{\pm}$.
On expanding the fermion operators in terms of
the eigenstates $\psi_E$
of the Dirac Hamiltonian $H_+ = \gamma^{0}  \gamma^{i}
(p^{i} - e A^{i}) \ + \ m \gamma^{0} $ (in the $A_0 =0$ gauge),
symmetrising with respect to charge-conjugation, and
subtracting the infinite vacuum contribution,
 one obtains
\bea
{1 \over 2} C_{+}(x;m\to 0)
 &= & -{1 \over 4} {\lim_{m \to 0}} \sum_{E} \mbox{sign}(E)
\psi_{E}^{\dag} \sigma^{3} \psi_{E} \ |^{F}_{0} \; .
\label{spin2}
\eea

The summation in (\ref{spin2}) symbolically denotes a sum over
discrete states and an integral over the continuum.
In the $m \to 0$ limit, the $E >0$ and $E<0$ eigenstates are
related \cite{ACJ}
by $\psi_{-E} = \sigma^{3} \psi_{E}$, so that
only the zero modes contribute to (\ref{spin2}).
For $F>0$ the zero modes occur at $E=m \to 0$ and are of the form
$\psi_{0} \sim \pmatrix{u \cr 0}$,
while for $F<0$ they occur at $E=-m \to 0$ and are of the form
$\psi_{0} \sim \pmatrix{0 \cr v}$.
Denote by $I$ the smallest integer greater
than or equal to $|F|-1$; then for a given $m=0^{\pm}$,
there are $I$ discrete (normalisable
to unity) states
in addition to the continuum (scattering) states \cite{ACJ,BB}.
Eq.(\ref{spin2}) therefore becomes
\begin{eqnarray}
{1 \over 2} \ C_{+}(x;m\to 0^{\pm}) &=& -{ \mbox{sign}(m)
\over  4} \sum_{E=m \to 0^{\pm}}
\psi_{E}^{\dag}  \psi_{E} |^{F}_{0} \;\, . \;\;\;\;\;
\,  \label{zer}
\end{eqnarray}
The discrete zero modes in the sum (\ref{zer})
are localised \cite{ACJ,BB} around the magnetic field and so
give a nonzero local contribution. On the other hand the
continuum zero modes, being scattering states, give
a negligible local contribution \cite{Pol} unless the flux is infinite.
However the continuum states are important to establish the
index theorem which is obtained by integrating (\ref{zer}) over all space :
The discrete states  contribute an amount
$- \ {\mbox{sign}(m) \over 4} \ I $, while a careful analysis
\cite{Pol}
shows that the scattering
 states contribute $- \ {\mbox{sign}(m) \over 4} \ (|F|-I)$,
\begin{eqnarray}
{\cal{S}} &\equiv& \int d^2 x { C_{+}(x;m \to 0^{\pm}) \over 2} =
-{\mbox{sign}(m) \over 4} \ | F | \, \label{spin3} .
\end{eqnarray}

The discussion above has been for regular field configurations.
If the flux  contracts to a single
point to form an AB flux string then
no  discrete normalisable states form even for $|F| > 1$ because
of the highly singular nature of the field. Rather,
if one starts
from a finite size tube and shrinks it to zero,
 the would be discrete
states collapse to point-like states located on the string.
These hidden states maintain the relations (\ref{c3}, \ref{ang},
\ref{spin}) in the AB case when $|F| >1$ but then they cause
a breakdown in
periodicity, with respect to
the flux, of induced quantum numbers in that configuration
\cite{PG,FL}.

\subsection{The AB case}

The purpose here is to verify (\ref{spin3}) explicitly for an AB string
with $|F| <1$.
The eigenstates of the AB hamiltonian in polar coordinates, and
in the $A_0 =0, \, e \ \vec{A}
= {F \over r} \ \hat{\theta} $ gauge, are readily determined,
\bea
\psi_{E,n} &=& \left[(E-m)k \over 4 \pi E \right]^{{1 \over 2}}
\pmatrix{ {k \ e^{i(n-1)\theta} \over E-m} \
\epsilon_{n} \ J_{\epsilon_{n}(l-1)}
(kr) \cr e^{i n \theta} \ J_{\epsilon_{n} l}(kr)} \, , \label{wf}
\eea
\bea
l \equiv  n-F, &\;\;\;&  n=\mbox{integer}, \
|F| <1, \ F \neq 0 \ , \nonumber \\
&& \nonumber \\
E = \pm \sqrt{k^2 + m^2}, \ && 0 \le k \le \infty , \,\, m \neq
0 \ , \nonumber \\
&& \nonumber \\
\epsilon_n = +1  \ \mbox{for} \ n \ge 1 \;\; & \mbox{and} & \;\;
\epsilon_{n} = -1 \ \mbox{for} \ n \le 0 \, . \nonumber
\eea
In the above, the sign of $\epsilon_n$ has been determined by
the usual requirement of square-normalisability \cite{Ger}
of the wavefunctions
and a possible
ambiguity for the critical partial wave $0<l_c < 1$ is
resolved by considering the string to be a limiting case of a
flux tube of finite size \cite{AB}. Notice that the critical
partial wave spinor contains one component which is singular
at the origin. Also note that the values $F=0, \pm 1$, which have been
excluded from (\ref{wf}) for technical convenience,
can be obtained in matrix elements (the induced charge,
spin etc.) by taking the limits $F \to 0, \pm 1$.

Substituting (\ref{wf}) into the right-hand-side of (\ref{spin2})
and integrating over all space gives
\bea
{\cal{S}} &=& {- 1 \over 8 \pi} \int d^2 x \  \left( m
\sum_{n} \int_{0}^{\infty} {dk \ k \over \sqrt{k^2 + m^2}}
\left\{ J^{2}_{\epsilon_{n} (l-1)} + J^{2}_{\epsilon_{n}(l)}
\right\}|_{0}^{F} \right)_{m \to 0} \  \label{ABspin}
\eea
where the massless limit has to be taken { \it inside} the space-integral.
Before evaluating this quantity, consider first the same
expression but with the massless limit taken  after doing
the spatial integral. Calling this last object ${\cal{T}}_{m \to 0}^{AB}$,
one has
\bea
{\cal{T}}^{AB} &=&
 -{m \over 8 \pi} \int d^2x \int_{0}^{\infty}
{dk k \over \sqrt{k^2 + m^2} } \sum_{n \ge 0} G_n (kr) \left|_{0}^{F}
\right.
\eea
where
\bea
G_n(kr) &\equiv & J_{n-F}^{2}(kr) + J_{n+F}^{2}(kr) + J_{n-F+1}^{2}(kr)
+ J_{n+F+1}^{2}(kr) \, .
\eea
The quantity ${\cal{T}}^{AB}$ is discussed at length in Appendix A.

Let us now return to the evaluation of (\ref{ABspin}),
that is, of $\cal{S}$. For this we need the massless limit of the
wavefunctions in (\ref{wf}). It is easy to verify that for $m \equiv 0$,
$\psi_{-k,n}
= - \psi_{k,n}$ and if this is used indiscrimately in (\ref{spin2}) one
obtains zero
for the induced spin. The resolution of this paradox
lies in recalling that there ought to be
unpaired continuum zero modes ($E = 0^{\pm}$). In
order to isolate these unambiguously
one has to start with $m \neq 0$.
Now when $m \neq 0$ (that is before the massless limit),
the states with energy $E \sim \pm m$ correspond to values
of $k$ satisfying $k \ll |m|$. Thus one considers the
${ k \over  |m|} \to 0$ limit in (\ref{wf})
to identify the nonvanishing threshold states.
In this limit only the critical partial wave survives:
For $ 1>F >0 $ its limiting form is
\bea
\ \lim_{k \to 0} \;  \Psi_{n_c =1 }  & =& \mbox{sign}(m)
\ \sqrt{k \over 2 \pi}
\pmatrix{ J_{-F} (kr) \cr 0 }, \;\, E \approx m , \;   F>0 \, ,
\label{s1}
\eea
while for $0 > F> -1$ it is
\bea
\ \lim_{k \to 0} \;  \Psi_{n_c =0 }  & =&
\ \sqrt{k \over 2 \pi}
\pmatrix{ 0 \cr J_{F} (kr) }, \;\, E \approx -m , \;  F<0 \, .
\label{s2}
\eea
(The Bessel function $J_{|F|}(kr) $ has not been expanded in $k$ because
$kr$ can be large even if $k$ is small).
When finally the limit $m \to 0$ is taken, these (\ref{s1}-\ref{s2})
will be the only
states which will contribute to the spin, the contribution from the
$|E| > 0$ states cancelling in pairs as mentioned above.
Thus for $ |F| <1$ the \cite{com}
scattering zero modes (\ref{s1}-\ref{s2})
contribute to (\ref{spin2}) the amount
\bea
{1 \over 2} \langle {\bar{\Psi}}_{+} \Psi_{+} \rangle_{m \sim 0} &=&
-{\mbox{sign}(m) \over 4} \int_{0}^{\lambda} {dk \over 2 \pi} \ k \
J_{-|F|}^{2}(kr) |^{F}_{0} \\
&& \nonumber \\
&=& - {\lambda^2 \mbox{sign}(m) \over 16 \pi} \left\{ J_{-|F|}^2(\lambda r) -
J_{-|F|-1} (\lambda r) \ J_{-|F|+1}(\lambda r) \right\} |^{F}_{0} \, ,
\label{ABden}
\eea
where  $\lambda \ll |m|$.
Note that taking the limit $|m| \to 0$ (and so $\lambda \to 0$)
 on the right-hand-side of
(\ref{ABden}) results in a null contribution to the (local) spin-density
in accordance with the remarks in the previous subsection that
for $|F| <1$ only the dilute continuum zero modes contribute to
the density.
However the space-integral of the density in (\ref{ABden}) is
non-negligible and is given by
\bea
2 \pi \ \mbox{sign}(m) \ \int_{0}^{\infty} { dr \over 2}  r
\langle {\bar{\Psi}}_{+} \Psi_{+} \rangle_{m \sim 0}
&=& -{\lambda^2 \delta(0) \over 8 \lambda} \ |^{F}_{0} + {\lambda^2 \over 8}
\int_{0}^{\infty} dr\ r \left\{ J_{-|F|-1}(kr) \ J_{-|F|+1}(kr)
\right\}|^{F}_{0} \nonumber \\
&& \nonumber \\
&=& {1 \over 8} \int_{0}^{\infty} dx \ x \left\{
J_{-|F|-1}(x) \ J_{-|F|+1}(x) \right\}|^{F}_{0} \label{int1} \\
&=& {1 \over 8} (-2|F|) \nonumber \\
&=& - {|F| \over 4} \, , \nonumber
\eea
and so
\bea
2 \pi \int_{0}^{\infty} dr \ { r \over 2}
\langle {\bar{\Psi}}_{+} \Psi_{+} \rangle_{m \to 0} &=& - \ \mbox{sign}(m) \
{|F| \over 4} , \;\;\;\;\;
|F| <1 \, , \label{res}
\eea
which agress with the index theorem (\ref{spin}).
Note that the integral in (\ref{int1}) is independent
of the parameter $\lambda$
 so that the final
result is not sensitive to the slight ambiguity in the choice of this
cutoff. The evaluation of the integral (\ref{int1}) is
described in Appendix B.

To summarise the results of this section, the induced spin in the AB system
was computed for $|F| <1$ and the index formula (\ref{spin})
verified. It was shown that
 isolating the contribution of the threshold states required
 a careful consideration of the massless limit.
For $ |F|<1$ the contribution to $\cal{S}$ is smeared over all space
due to the
nonlocalised nature of the continuum  (scattering) threshold states
and thus the spin-density (in the required massless limit) is zero.  For
$|F| >1$ there
are also discrete states for any finite size flux tube
which give a localised nonzero contribution to $\cal{S}$
(see the analogous discussion for the charge in \cite{PG,FL}).
As mentioned earlier, these discrete states
collapse to point-like states in the limit of a zero-radius flux
tube.

The presence of induced quantum numbers around the $QED_3$
AB string implies that it is detectable (through a
scattering experiment) even for integral flux \cite{PG}.
However, there is a return flux at infinity (actually, at
$r \sim 1/ \mu $ where $ \mu \sim e^2$ is the dynamically
generated photon mass)  complementary to the
AB flux in the string, and one can ask if the vacuum
polarisation effects associated with that return flux
can aid in restoring the invisibility of the integral flux string.
If the mass $m$ of the electron is much larger than
$\mu$, then for any $F$
the polarisation effects (\ref{c3}-\ref{ang}) are localised within
$1/m$ of the string and are probably not significantly \cite{PG}
affected  by the return flux
effects located at $1/\mu$ .
For $F$ exactly an integer the induced
effects are exactly on the string and so should be
even more immune from the
return flux effects.

The situation is more complicated
for massless electrons, which is the limit in which the
induced spin (\ref{spin})
is defined. However even in the massless
case, the contribution from the collapsed threshold states (for
$|F| >1$ ) is highly localised so it is not very clear to me
if these
bound states will be completely nullified in their contribution
to (\ref{c3}-\ref{spin}) by feedback effects.
One does however expect the contribution  from the scattering
states to be cancelled by feedback
\cite{LCG}.

There are other subtleties associated with the massless
theory.
For example,
the factor of $\mbox{sign}(m)$ in (\ref{res}), if averaged over,
 suggests
no polarisation effects in the massless limit.
However if one adopts the point of view
that the theory is sensibly defined only for massive electrons
(to define an unambiguos zero of energy)
then no matter how small that mass is with respect to $e^2$,
the $\mbox{sign}(m)$ factor distinguishes between the two {\it inequivalent}
physical representations
one is working in and so should not be averaged over.
Indeed, even if one starts with $m \equiv 0$ in  (\ref{lag2}),
a small nonperturbative mass is generated dynamically
\cite{chiral}. This suggests that the ambiguities of a truly massless
theory are probably academic and one should at most speak
of a massless {\it limit}, where the fermion mass is very much
smaller than the other dimensionfull parameter, $e^2$.

The discussion in this section has been for the parity {\it non-}invariant
lagrangian (\ref{lag1}). In the parity-invariant theory (\ref{lag2}-
\ref{lag3})
there is no photon mass generation due to cancellations between the
contributions of the
"+" and "-" spinors, and so no induced charge like (\ref{c3}).

\setcounter{equation}{0}
\section{Chiral symmetry breaking}

Let $q$ be the Dirac fermion operator in four space-time
dimensions. Then to study
dynamical chiral symmetry breaking, one looks at
the condensate $<\bar{q} q >$ in the massless limit of the full dynamical
theory.
In general this is  difficult to do analytically and an intermediate
step that can give some insight is to look at the condensate for a
particular background gauge-field. In Ref.\cite{BC} it was shown that if
$<\bar{q} q >^{A}$ is the space-time average of the
condensate in the background field $A_\mu$,
and $\rho_{D}(\lambda)$ is the density of eigenvalues of the background
operator
$i \ \slash$ \hspace{-4.5mm}$D$ per unit space-time volume, then
\bea
<{\bar{q}} q>^{A} &=& \pi \rho_{D}(0) \, . \label{BC}
\eea
That is, chiral symmetry breaking in a given background field is related to
the spectral density at zero energy of the operator
$i \ \slash$ \hspace{-4.5mm}$D$.
If one can
parametrize the right-hand-side of $(\ref{BC})$ explicitly in terms of the
background field $A_\mu$, then by integrating (in the path-integral)
Eq.(\ref{BC}) over all possible
gauge-field configurations, one can in principle study
dynamical chiral symmetry
breaking in the full theory.
In practice however, it is already quite difficult to
achieve the first step above :
that is, to make  statements about
(\ref{BC}) for arbitrary configurations.

However for three-dimensional QED described by
the parity-invariant lagrangian (\ref{lag2}), much can be said.
For external time-independent magnetic fields with flux $F$,
the condensate is given by (\ref{cond2}-\ref{zer}) to be
\begin{eqnarray}
 C(x;m\to 0^{\pm}) &=& - \mbox{sign}(m)
 \sum_{E=m \to 0^{\pm}}
\psi_{E}^{\dag}  \psi_{E} |^{F}_{0} \;\, ,  \label{zer2}
\end{eqnarray}
where $\psi_E$ are the eigenstates discussed in the previous section.
It is quite clear that (\ref{zer2})
is just a special case of (\ref{BC}), obtained by restricting
to time-independent magnetic fields. However (\ref{zer2})  contains some
unique features which enhance its
usefulness in spite of the restrictions.
For any finite flux, the
spectral density on the right-hand-side of (\ref{zer2}) is dominated
locally by a finite number of
discrete (normalisable to unity) zero modes. The number of such modes is
determined solely by the flux as discussed earlier.
Furthermore, the integrated density
also depends only on the flux,
\bea
 \int_{0}^{\infty} d^2 x
\langle {\bar{\Psi}}_{+} \Psi_{+} \rangle_{m \to 0} &=& - \ \mbox{sign}(m) \
|F| \ . \label{res2}
\eea
This last is significant for it means that for any
configuration with nonzero net flux, the chiral condensate
must be nonzero at some place.
Thus one can decide if chiral-symmetry is broken or not
by just knowing the flux rather than the details of the magnetic
field.
This also suggests that the contribution
of static magnetic fields to the full dynamical condensate
is not negligible (of nonzero measure in the
path-integral) and makes it plausible that dynamical
chiral symmetry happens in the full dynamical theory (that is, without
external fields). Of course
the last is already known to happen \cite{chiral}
and so the real challenge is to see if one can use
(\ref{zer2}-\ref{res2}), integrated over
arbitary static magnetic fields (of variable fluxes) to get an
estimate for the dynamical condensate. By ignoring
time-dependent fields, and in particular electric
fields, there is no {\it a priori}  reason such an  estimate would
be close to the results of Ref.\cite{chiral}, but still
the exercise  might be revealing if doable.
(This will not be attempted here.)

Another important feature of (\ref{zer2}) and (\ref{res2}) is the sign$(m)$
factor. As noted in Ref.\cite{RP} this
means that the direction of
chiral symmetry breaking in external magnetic
fields is correlated with the infinitesimal breaking of that symmetry
brought about by a mass $m = 0^{\pm}$ (see the discussion of massless
theories near the end of the last section).

Let me mention now some perturbations which could destroy the
condensate (\ref{zer2}-\ref{res2}) formed by static magnetic fields.
Viewing the condensate as an ordering of spins
in a magnetic field, one expects
a superimposed electric field to reduce such an ordering and so
cause a reduction in the magnitude of the condensate.
At least if the
electric field is weak then one might be able to
check this in a perturbative
treatment as was done
in  Ref.\cite{MM} for a Chern-Simons theory.

Thermal fluctuations are also naturally expected to
disrupt any ordering (symmetry breaking). For example,
the charge (\ref{c3}) has been computed \cite{rev,N}
for slowly varying magnetic fields at nonzero temperature
and density, showing a rapid "washing out" at high temperature.
More remarkably, for any non-zero temperature, the charge
was shown to vanish in the massless limit; a simple explanation of
this given in \cite{rev}
was that in the massless theory the charge is due to the zero-energy modes
which are susceptible to large fluctuations in their occupancy at nonzero
temperature. Since the chiral condensates ({\ref{zer2}-\ref{res2}) also
receive support only from the zero modes,
the qualitative argument given
above suggests that they should also vanish at nonzero temperature.
This has recently been noted in an explicit calculation
for a constant magnetic field \cite{DH}.

\setcounter{equation}{0}
\section{Conclusion}

In this paper were studied some phenomena in three-dimensional QED in an
external magnetic field.
The breaking of chiral symmetry
by  an external
magnetic field in the reducible representatiuon
of QED$_{3}$  (\ref{lag2})
is equivalent to the polarisation of spin in the irreducible
theory (\ref{lag1}). Thus chiral symmetry breaking in the toy-model
(\ref{lag2}) may be heuristically understood as an
ordering of spins in a magnetic field. Then any effect which interferes
with this ordering, such as an electric field or thermal fluctuations,
would be expected to diminish the breaking of
chiral symmetry.

Quantitatively, the chiral condensate in an external magnetic field is
controlled by the total flux of that field (\ref{zer2})
as it determines the number
and nature (discrete and localised {\it versus} continuum and non-localised)
of the zero modes of the massless Dirac equation. For any finite flux,
the localised zero modes give the essential non-negligible contribution
to the chiral-condensate.
The space-integral of the
condensate receives a nontrivial contribution also from the
dilute scattering zero modes and the net result
is an invariant depending only on the flux (\ref{res2}). This
index formula
gives the phenomena a certian robustness and  might
be useful in obtaining estimates of the chiral condensate in the full
dynamical theory as described in Sect.3.

The polarisation effects (\ref{c3}, \ref{ang}, \ref{spin}) in the
theory (\ref{lag1}) are also invariants depending only on the
flux. It has
been verified (\cite{others,PG,FL,cone} and Sect.2)
that the formulae (\ref{c3}, \ref{ang}, \ref{spin})
hold also for singular
configurations such as the AB flux string. Consequently,
as a result of the localised
threshold states which form every time the flux passses an integer, there
accumulate around an AB string vacuum polarisation effects which are not
periodic in the flux. This exposes an integral flux ($F \neq 0$)
string to external probes \cite{PG}. By contrast, such polarisation
effects cancel for flux tubes in
QED$_4$ due to the four component
nature of the fermions which preserve parity \cite{current,FL}.
However this does not rule out other polarisation effects
(as yet undiscovered)
which might also make an integral AB flux string in QED$_4$
visible.

In the model (\ref{lag2}) the net induced charge around a magnetic field
vanishes because the contribution from the $"+"$ and $"-"$ two-component
spinors cancel (since they have opposite signs of mass). One can consider
a modification of (\ref{lag2}) which is also parity invariant but in which the
net induced charge is nonzero. This model is obtained by replacing the
charge $e$ in (\ref{lag2}) by the matrix $e^{\star} =
e \ \pmatrix{I &0 \cr 0 & -I}$.
This means that the $\pm$ spinors now have opposite charges in addition to
oppositely signed masses and their respective contribution to the
induced charge
will be additive. This last model
has actually been used as a parity-invariant effective lagrangian for
high-temperature superconductivity,  but the gauge-field which couples to
the charge $e^{\star}$ there is a
"statistical gauge field" \cite{tc}
rather than the true electromagnetic field.

There certainly exist  vacuum polarisation effects around flux tubes,
even in parity-odd three-dimensional QED (\ref{lag1}),
which are not caused by parity breaking effects like fermion
masses. For example,
there are vacuum
currents which circulate around thin magnetic flux tubes and are
due to the $AB$ effect acting on virtual electron-positron
pairs \cite{current,PG,FL}.
These currents occur also for scalar theories and in
 $3+1$ dimensions.
A possible role for them  has been proposed recently in a picture
of a nonperturbative vacuum of QED$_4$ \cite{WS,GLP}.

Whether one uses three-dimensional QED as a simplified model in particle
physics, as an effective theory in condensed matter physics, or as
a field theory of intrinsic interest, it is clear that much
more remains to be understood. The focus in this paper has been on
extending previous discussions of polarisation effects due to magnetic
fields. The interested reader is referred to some other recent works
\cite{more} in which
magnetic fields are also employed to unravel the dynamics of
QED$_3$.\\ \\

{\bf Acknowledgements}\\
I thank  Alfred Goldhaber and Hsiang-nan Li for helpful comments, and Pijush
Ghosh for Ref.\cite{MM}.

\newpage

\renewcommand{\theequation}{A.\arabic{equation}}
\setcounter{equation}{0}
\noindent
\noindent {\large\bf Appendix A}
\vskip 3mm \noindent

Here the quantity ${\cal{T}}^{AB}$ mentioned in Sect.2 is evaluated,
and some subtleties involved are discussed by comparing the calculation
with the corresponding one for the induced charge $\cal{Q}$.

Simplifying $\langle {\bar{\Psi}}_{+} \Psi_{+} \rangle_{+}$ for the AB case
one obtains,
\bea
\langle {\bar{\Psi}}_{+} \Psi_{+} \rangle_{+} &=&
{- 1 \over 4 \pi} \  \left( m
\sum_{n} \int_{0}^{\infty} {dk \ k \over \sqrt{k^2 + m^2}}
\left\{ J^{2}_{\epsilon_{n} (l-1)} + J^{2}_{\epsilon_{n}(l)}
\right\}|_{0}^{F} \right) \  \label{a0} \\
&& \nonumber \\
&=& -{m \over 4 \pi} \int_{0}^{\infty}
{dk k \over \sqrt{k^2 + m^2} } \sum_{n \ge 0} G_n  |_{0}^{F} \label{a1}
\eea
where
\bea
G_n(kr) &\equiv & J_{n-F}^{2}(kr) + J_{n+F}^{2}(kr) + J_{n-F+1}^{2}(kr)
+ J_{n+F+1}^{2}(kr) \, .
\eea
Note the manifest C-symmetry, $F \to -F$, of the expression for $G_n$.
By writing
\bea
{ 1 \over \sqrt{k^2 + m^2} } = {2 \over \sqrt{\pi} } \int_{0}^{\infty}
d\alpha \
 e^{-\alpha^2 (k^2 +m ^2)}, \label{trick}
 \eea
the $k-$integral in (\ref{a1}) may be performed using \cite{GR}
\bea
\int_{0}^{\infty} e^{-\alpha^2 k^2} J_{\mu}^{2}(kr) \ k \ dk
&=& {1 \over 2\alpha^2} \ e^{- {r^2 \over 2 \alpha^2}} \
I_{\mu} \left({r^2 \over 2 \alpha^2}\right) \label{gr1} \, ,
\eea
to give
\bea
\langle {\bar{\Psi}}_{+} \Psi_{+} \rangle_{+}
&=& -{m \over 4 \pi \sqrt{\pi} }
\int_{0}^{\infty} {d\alpha  \over \alpha^2}  \
e^{-\alpha^2 m^2 -z} \left. \left( \sum_{n \ge 0}
H_{n}(z)  \right) \right|_{0}^{F}  \label{a2}
\eea
where
\bea
H_{n}(z) &\equiv& I_{n-F}(z) + I_{n+F}(z) + I_{n-F+1}(z) + I_{n+F+1}(z)
\eea
and $ z = r^2 / 2 \alpha^2$. Now consider the spatial $\int r dr d \theta$
integral over the density in (\ref{a2}).
The $\alpha-$integral can then be done
explicitly because it reduces to the simple Gaussian
\bea
\int_{0}^{\infty} d \alpha \ e^{- \alpha^2 m^2} &=& {\sqrt{\pi} \over 2 |m|}
\,
\eea
after the change
of variable $ r^2 =2\alpha^2 t$ in the $r-$integral. Thus one is left with
\bea
\int d^2 x \langle {\bar{\Psi}}_{+} \Psi_{+} \rangle_{+} &=& -{\mbox{sign}(m)
 \over 4 }
\int_{0}^{\infty} {dt}
e^{-t} \left. \left( \sum_{n \ge 0} H_{n}(t)  \right) \right|_{0}^{F}
 \label{a3} \, .
\eea
Assuming that the order of integration in (\ref{a3}) may be interchanged
with the summation, the $t-$integral may be evaluated by taking appropriate
limits of the formula \cite{GR}
\bea
\int_{0}^{\infty} e^{-at} I_{p+\nu} (bt) &=& {b^{p+\nu} \over
\sqrt{a^2 -b^2} (a+ \sqrt{a^2 - b^2} )^{p+\nu} }.
\eea
For example, for any integer $n$,
\bea
\int_{0}^{\infty} e^{-t} ( I_{n +F}(t) - I_{n}(t))  &=& -F \, . \label{tint}
\eea
Thus  the final result for the space-integral of (\ref{a0}) is
\bea
\int d^2 x \langle {\bar{\Psi}}_{+} \Psi_{+} \rangle_{+} &=&
-{\mbox{sign}(m) \over 4} \ \sum_{n \ge 0} \ (F-F+F-F) = 0  \, .
\label{tab}
\eea
Therefore ${\cal{T}}^{AB} = {1 \over 2} (\mbox{Eq.(\ref{tab})}) =0$,
 if the space-integral above may be performed
before the infinite partial wave summation.

By way of comparison consider the charge (\ref{c2}) evaluated
for the AB case with $|F| <1$. Substituing (\ref{wf}) into
(\ref{c2}) and simplifying gives \cite{PG},
\bea
{\cal{Q}} &=&
{- m e \over 4 \pi} \  \int d^2 x \left(
\sum_{n} \int_{0}^{\infty} {dk \ k \over \sqrt{k^2 + m^2}}
\left\{ J^{2}_{\epsilon_{n} (l-1)} - J^{2}_{\epsilon_{n}(l)}
\right\}\right) \  \label{ac0} \\
&& \nonumber \\
&=&
{- m e \ \mbox{sign}(F) \over 4 \pi} \  \int d^2 x \left(
\int_{0}^{\infty} {dk \ k \over \sqrt{k^2 + m^2}}
\left\{ J^{2}_{-|F|} - J^{2}_{|F|}
\right\}\right) \  \label{ac1} \\
&& \nonumber \\
&=&
- \mbox{sign}(m) { e F \over 2} \label{ac2}.
\eea
Apart from a factor  $2e$, the expression (\ref{ac0}) differs from
${\cal{T}}^{AB}$ by a relative minus sign between the two Bessel
functions in curly brackets. Consequently the sum over partial
waves in (\ref{ac0}) results in several cancellations to give (\ref{ac1}).
By writing $J_{|F|} - J_{-|F|} = (J_{|F|} -J_0 ) - (J_{-|F|} - J_0)$,
Eq.(\ref{ac1})
may be evaluated using (\ref{trick},\ref{gr1}, \ref{tint}) to get (\ref{ac2}).
(This computation is slightly different from the one in Ref.\cite{PG}
where (\ref{tint}) was not used).

Since $\cal{Q}$ is C-odd no $F=0$ subtraction is necessary in (\ref{ac0})
as that piece is automatically zero. On the other hand since $\cal{T}$
(and $\cal{S}$) are C-even, a $F=0$ subtraction is necessary to remove
an infinite bare vacuum contribution. The last infinity is due to the fact
that the spin of free fermions is given by $s = {1 \over 2} \ \mbox{sign}
(mE)$ so that the spin of the bare ground state,
$\sum_{E} \ \mbox{sign}(E) \ s \ \sim \ \sum \ \mbox
{sign}(m) $, is divergent.

If the inifinite summation over partial waves in (\ref{ac0}) is
regulated by a cut-off, $|n|_{\mbox{max}} = N$, ($N >0$)
 which is later taken to
infinity, one obtains (see \cite{LCG}),
\bea
{\cal{Q}} &=& {- m e \over 4 \pi} \  {\lim_{N \to \infty}} \
\int d^2 x \left(
\sum_{-N}^{N} \int_{0}^{\infty} {dk \ k \over \sqrt{k^2 + m^2}}
\left\{ J^{2}_{\epsilon_{n} (l-1)} - J^{2}_{\epsilon_{n}(l)}
\right\}\right) \  \label{ac20} \\
&& \nonumber \\
&=&
{- m e  \over 4 \pi} \   {\lim_{N \to \infty}} \
\int d^2 x
\int_{0}^{\infty} {dk \ k \over \sqrt{k^2 + m^2}}
\left( \left\{ J^{2}_{-F} - J^{2}_{F}
\right\} + \left\{ J^{2}_{N-F} - J^{2}_{N+1+F} \right\}
\right) \, . \label{ac21}
\eea
It is easy to verify using the formulae given earlier that if the
limit $N \to \infty$ in (\ref{ac21}) is taken {\it after} the space-integral,
the answer is zero as the contribution from the $N$-dependent Bessel functions
cancels the contribution from the first two. However, as noted in
Ref.\cite{LCG}, in the large $N$ limit the contribution from the $N$-
dependent pieces recedes to infinity and it may be interpreted
as representing the
oppositely charged polarisation cloud "repelled" by the
localised flux. So if one is interested in the local
polarisation effects, the $N \to \infty$ limit should be taken
before the space-integral in (\ref{ac21}). Note also that
for any finite $N$ the expression (\ref{ac21}) is no longer
C-odd, this asymmetry being broken by the $N$-dependent pieces.
This may be remedied by taking the C-odd part
of the truncated sum in (\ref{ac20}). The rest of the discussion in this
paragraph still goes through then.

Returning now to the evaluation of ${\cal{T}}^{AB}$, consider
regulating the partial-wave sum (\ref{a3}) in a $C$-even
way as for the charge above. Clearly, the result is still zero if the
space-integral is done first  before the summation.
This suggests, by analogy with the charge case above,
that in this order (space-integral before summation)
of calculating $\cal{T}$ one is adding the local
polarisation effects to the opposite non-local ones. Unfortunately,
because there are
no pair-wise cancellations among the partial waves in (\ref{a3})
as in the charge case, it is not obvious which
of the partial waves are contributing to the local polarisation effects
and which pieces may be interpreted as the opposite effects which
recede to infinity in the $N \to \infty$ limit.
Fortunately however this question, though interesting, is not of direct
relevance because the physical object $\cal{S}$ as defined in (\ref{spin3})
differs from $\cal{T}$ in that the massless limit has to be taken
before the space-integral (as suggested by the general
derivation sketched in Sect.2 ). The evaluation of $\cal{S}$ is described
in the main text.\\

\newpage

\renewcommand{\theequation}{B.\arabic{equation}}
\setcounter{equation}{0}
\noindent
\noindent {\large\bf Appendix B}
\vskip 2mm \noindent

Here the integral in Eq.(\ref{int1}) is evaluated. It  is
\bea
&&\int_{0}^{\infty} dx \ x \left(
J_{-|F|-1}(x) \ J_{-|F|+1}(x) -  J_{-1}(x) \ J_{1}(x)
\right) \nonumber \\
&&\;\;\;\;\;\;\;\;\; =  \int_{0}^{\infty} dx \ x \left(
J_{-|F|-1}(x) \ J_{-|F|+1}(x) + J_{1}^{2}(x)
\right) \label{bint2}
\eea
with $|F| <1$. The integral (\ref{bint2}) is well defined but
inconvenient to evaluate as a whole. Therefore the integral over the
$F$-dependent and $F$-independent Bessel functions in (\ref{bint2}) will
be performed separately. This will require an intermediate regularisation.
The method adopted here is the following. One has an integral \cite{GR}
\bea
\int_{0}^{\infty} J_{\mu}(t) \ J_{\nu}(t) \ t^{-\alpha} dt &=&
{ \Gamma(\alpha) \ \Gamma({ \nu + \mu - \alpha +1 \over 2})
\over 2^{\alpha} \Gamma({-\nu + \mu + \alpha +1 \over 2}) \
\Gamma({\nu + \mu + \alpha + 1 \over 2}) \ \Gamma({ \nu -\mu + \alpha +1
\over 2}) } \, , \label{balt} \\
&& \nonumber \\
\mbox{with} && \nu + \mu + 1 > \alpha > 0 \, . \label{blim}
\eea
By setting $\mu = -|F|-1$ and $\nu = -|F| +1$ in (\ref{balt}) and choosing
$F$ and $\alpha$ so that the condition (\ref{blim}) is satisfied,
the $F$ dependent integral in (\ref{bint2}) may be performed
by  analytically continuing the right-hand-side of (\ref{balt})
to $\alpha = -1$ and the full range of $F$. The $F$-independent
integral in (\ref{bint2}) may be evaluated similarly. The net result for
(\ref{bint2}) is $-2|F|$.

It is useful to have a cross-check on the above manipulations
with (\ref{balt}). For this purpose, consider the following
calculation of the charge (\ref{c2}), for the AB system with
flux $|F| <1 $, in the massless limit of the theory.
Using the
continuum threshold states given by (\ref{s1}
-\ref{s2}) and the mapping $\psi_E = \sigma^3 \psi_{E}$ between the
non-threshold states, one obtains
(cf. the calculation of the spin in \ref{ABden}-\ref{res})
\bea
{\lim_{m \to 0}} \
 {\cal{Q}} &=& { -e \over 2} \mbox{sign}(m) \ \mbox{sign}(F) \
{\lim_{\lambda \to 0}}
\ \int_{0}^{\infty} 2\pi r dr \ \int_{0}^{\lambda} { dk \over 2 \pi}
\ J_{-|F|}(kr) \ |_{0}^{F} \label{bint3} \\
&& \nonumber \\
&=& { -e \over 2} \mbox{sign}(m) \ \mbox{sign}(F) \ |F| \\
&& \nonumber \\
&=& { -e \over 2} \mbox{sign}(m) \ F \ , \label{bint4}
\eea
where $\lambda \ll |m|$.
In (\ref{bint3}) a $F=0$ subtraction has been inserted to remove
divergent $F$-independent pieces. Such a subtraction was not
required in the calculation of $Q$ (for $m \neq 0$)
by the conventional method
in Appendix A (see \ref{ac0}) and so presumably it is required here
because of the ambiguity in the cut-off $\lambda$.
The integral in (\ref{bint3}) has been evaluated using (\ref{balt})
and the final result (\ref{bint4}) agrees with the expected formula
(\ref{c3}) which holds independent of the magnitude of $m$.
(Please note however that the charge-{\it density} is not a
topological invariant and it also depends nontrivially on the
mass $m$. Thus, for example, for $|F| <1$ the charge-density is infinitely
dilute in the $m \to 0$ limit (receiving contributions only from continuum
zero modes (\ref{bint3}))
but is concentrated within a distance $1/m$ of the localised field
for $m \neq 0$ (in which case it receives contributions also from
non-threshold continuum states) \cite{PG,FL}.)



\begin{thebibliography}{99}
\bibitem{com}
Actually because of the normalisation
factor $\sim \sqrt{k}$ the scattering zero modes
in (\ref{s1}-\ref{s2}) are nonvanishing as $k \to 0$ only for
$|F| > 1/2$ if $r$ is held fixed. However if $r \to 0$ much
faster than
$k \to 0$ then these states are nonvanishing for the full range
$|F| <1$.
\end{thebibliography}
\end{document}